\documentclass[proof]{WileyASNA-v1}

\articletype{Article Type}%

\received{}
\revised{}
\accepted{}

\begin{document}

\def\gtsima{$\; \buildrel > \over \sim \;$}
\def\ltsima{$\; \buildrel < \over \sim \;$}
\def\gsim{\lower.5ex\hbox{\gtsima}}
\def\lsim{\lower.5ex\hbox{\ltsima}}

\title{Observations of X-ray reverberation around black holes}

\author[1]{Barbara De Marco*}
\author[2,3]{Gabriele Ponti}


\authormark{B. DE MARCO \textsc{et al}}

\address[1]{\orgname{Nicolaus Copernicus Astronomical Center, Polish Academy of Science}, \orgaddress{\state{Bartycka 18, PL 00-716, Warsaw}, \country{Poland}}}
\address[2]{\orgdiv{Max-Planck-Institut f\"ur Extraterrestrische Physik}, \orgaddress{\state{Giessenbachstrasse, D-85748 Garching}, \country{Germany}}}
\address[3]{\orgdiv{INAF-Osservatorio Astronomico di Brera}, \orgaddress{\state{Via E. Bianchi 46, I-23807 Merate (LC)}, \country{Italy}}}

\corres{*B. De Marco \email{bdemarco@camk.edu.pl}}

\presentaddress{\orgname{N. Copernicus Astronomical Center, PAN}, \orgaddress{\state{Bartycka 18, PL 00-716, Warsaw}, \country{Poland}}}

\abstract{The X-ray emission from accreting black hole (BH) systems displays strong variability. Short reverberation lags are expected between the primary hard X-ray continuum and the reprocessed disc emission. These lags depend on light-travel distances, thus offering the opportunity to map the geometry of the innermost accretion flow. X-ray reverberation lags have been observed in several BH accreting systems. In radio quiet active galactic nuclei (AGN) these lags scale with BH mass and point to a reprocessing region located close to the Comptonizing X-ray corona. On the other hand, reverberation lags detected in the hard state of some BH X-ray binaries (BHXRB) suggest a different accretion flow geometry than in AGN, showing evidence of evolution as a function of luminosity. }




\maketitle


\section{Introduction}\label{sec:intro}

Accretion of gas onto a black hole (BH) is a fundamental source of energy extraction in the Universe, and the only mechanism able to explain the existence of some of the most powerful known astrophysical sources. These include active galactic nuclei, AGN, black hole X-ray binary systems, BHXRB, {\it some} ultra-luminous X-ray sources, ULXs, and tidal disruption events, TDEs.\\
Early observations of AGN and BHXRBs (e.g. Lightman \& Shapiro 1975; Nandra \& Pounds 1994) highlighted the presence of at least two main zones of energy dissipation close to the BH: an optically thick disc of relatively cold accreting gas and an optically thin phase of hot ionized plasma (e.g. Shakura \& Sunyaev 1973; Thorne \& Price 1975; Shapiro et al. 1976; Sunyaev \& Titarchuk 1980; Zdziarski et al. 1997; S\c{a}dowski et al. 2016). The latter constitutes the region where the thermal photons emitted by the disc are Compton up-scattered to hard X-ray energies. Moreover, under certain physical conditions (e.g. large scale-height), it might be related to the production of powerful jets (Meier et al. 2001).

Some of the outstanding questions regarding BH-accretion concern the geometry of the inner accretion flow: how is the hot Comptonizing gas spatially distributed? Does the accretion disc reach the innermost stable circular orbit (ISCO)? Does the distribution of the hot/cold phase of the inner flow evolve as a function of the accretion regime of the source and how?

Answers to these questions may yield the key to understand the complex and diverse phenomenology characterizing BH-accreting systems. Such phenomenology includes the existence of AGN accreting at different rates (e.g. Heckman \& Best 2014) and characterized by the presence/absence of a strong relativistic jet (e.g. Padovani et al. 2016); the spectral state changes observed in BHXRBs (e.g. Fender et al. 2004; Homan \& Belloni 2005; Ponti et al. 2012a), associated with a remarkable evolution of the broad-band X-ray variability characteristics (e.g. Belloni et al. 2005; Mu\~noz-Darias et al. 2011) and the occasional appearance of quasi periodic oscillations (QPO, e.g. Casella et al. 2005). 

In recent years, the increasing availability of data from large effective area X-ray detectors with good timing capabilities and the use of powerful {\it spectral-timing} Fourier techniques (Uttley et al. 2014) have been providing us new tools to address these questions. 
Some of the most relevant advances are in the study of {\it X-ray reverberation}. This investigates the time-delayed response of the inner disc to illumination of variable hard X-ray photons. The lag of the reprocessed component (produced via reflection and thermal heating, e.g. Guilbert \& Rees 1988) maps the physical distance between the hard X-ray source and the disc reprocessing region. Therefore, it is a powerful diagnostic of the geometry of the innermost accretion flow (e.g. Wilkins \& Fabian 2013; Uttley et al. 2014). 

X-ray reverberation has now been observed in AGN and BHXRBs (see following sections), as well as in a TDE (Kara et al. 2016a) and, possibly, in one ULX (Heil \& Vaughan 2010; De Marco et al. 2013a, Hern\'andez--Garc\'ia et al. 2015). In this contribution we will focus on observations of X-ray reverberation in AGN and BHXRBs, illustrating recent advances and future perspectives in the field.

\section{X-ray reverberation in AGN}\label{sec:revAGN}

Ten years after the first significant detection (Fabian et al. 2009), the observational evidences suggest that X-ray reverberation is a common feature of bright, radio-quiet AGN.
Long available {\it XMM-Newton} (and more recently NuSTAR) observations have, so far, allowed detection of lags ascribable to X-ray reverberation in over 30 sources. 

\subsection{Soft X-ray lags}
\begin{figure}
        \includegraphics[width=\columnwidth]{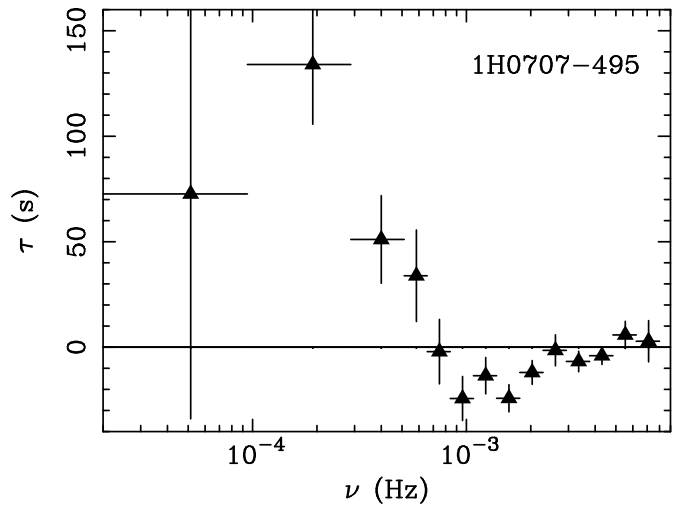}
    \caption{The lag vs frequency spectrum of 1H0707-495 between energy bands $E=0.3-1$ keV and $E=1-4$ keV. Negative lags indicate to the soft band lagging behind the hard band (from De Marco et al. 2013b).}
    \label{fig:1H}
\end{figure}

Early hints to the existence of the long-sought-for (e.g. Vaughan \& Edelson 2001) signature of X-ray reverberation were reported by McHardy et al. (2007) in the narrow line Seyfert 1 (NLSy1) galaxy AKN 564. However, the first clear and robust ($>5\sigma$) detection was obtained a few years later, from the study of a long ($>$500 ks) observation of the NLS1, 1H0707-495 (Fabian et al. 2009; Fig. \ref{fig:1H}).
In both cases, the lag was measured in the response of the soft excess (excess emission at $E\lsim$ 2 keV above the broad band X-ray continuum commonly seen in radio-quiet AGN) to short time scale variations of the primary hard X-ray power law, i.e. a so-called {\it soft lag}. While the power law component is thought to be produced in the hot ionized plasma located close to the BH (the putative {\it corona}, Fabian et al. 2017), there is much more uncertainty on the origin of the soft excess. The currently most popular models invoke either relativistically blurred reflection from an ionized accretion disc (Crummy et al. 2006; Fabian \& Ross 2010) or thermal Comptonization in a low temperature, optically thick corona (Done et al., 2012; Jin et al., 2013; Petrucci et al. 2018). The photoionized disc relativistic reflection solution was shown to provide a good fit to the spectra of 1H0707-495 (Zoghbi et al. 2010). Therefore, the detection of a short ($\sim$ 30 s), high-frequency ($\nu\gsim 10^{-3}$ Hz) soft lag (Fig. \ref{fig:1H}) associated with this component gave strong support to this interpretation. Indeed, the lag amplitude and its detection up to high temporal frequencies ($\sim 0.01$ Hz) suggest the reprocessing region to be very close (only a few gravitational radii, $r_g$) to a compact corona. This is expected if reprocessing occurs in the inner regions of a disc extending down to the ISCO.

This discovery triggered further investigation (e.g. De Marco et al. 2011; Emmanoulopoulos et al. 2011; Tripathi et al. 2011). As a matter of fact, X-ray reverberation off the inner disc should be detectable in all variable sources, not presenting strong obscuration or a relativistic jet. To address this question we carried out the first systematic study of X-ray reverberation in a large sample of local radio-quiet AGN (De Marco et al. 2013b), selecting the most variable, least obscured sources (chosen from the CAIXAvar sample of Ponti et al. 2012b), with sufficiently long XMM-Newton observations. 
\begin{figure}
        \includegraphics[width=\columnwidth]{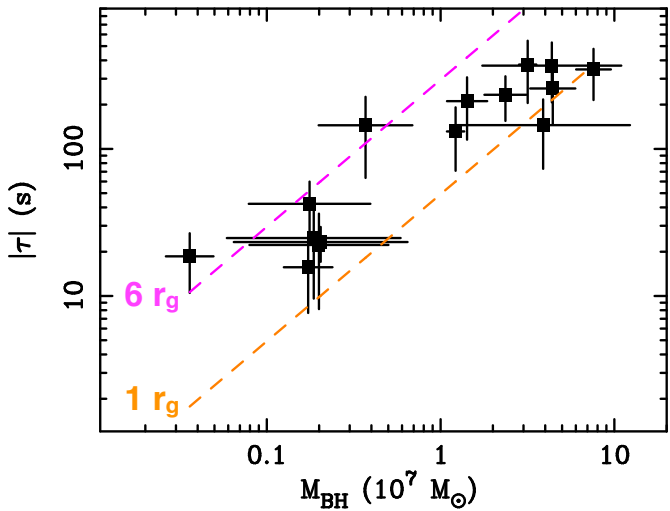}
    \caption{Soft X-ray reverberation lag amplitude vs $M_{BH}$. The dashed lines mark the light-crossing time of $1r_g$ and $6r_g$ as a function of BH mass (from De Marco et al. 2013b).}
    \label{fig:AGNlagMass}
\end{figure}
This study raised the number of detections of high-frequency soft lags to 15 (out of the 32 sources studied). Most importantly it revealed that in radio quiet AGN the lag amplitude and frequency scale approximately linearly with the BH mass\footnote{Published values of BH masses were used, primarily obtained from optical reverberation (e.g. Peterson et al. 2004), and in a few cases from single epoch methods (e.g. Zhang \& Wang 2006)} (Fig. \ref{fig:AGNlagMass}). Such a property suggests that the same reprocessing mechanism is responsible for the soft lag observed in different sources, and that all the systems have similar geometry. In agreement with results from 1H0707-495, the short distances mapped by the lags (in all the cases only a few $r_g$) as well as their frequencies, hint at the reprocessing region being very close to the ISCO and illuminated by a compact corona.

Finally, these results give independent clues about the origin of the soft excess. Indeed, the detection of a soft lag in potentially all the sources with good statistics on the relevant time scales\footnote{Sources with a marginal detection typically had $M_{BH}>10^7M_{\odot}$ and relatively short XMM observations. At such high masses the longer reverberation lag cannot be well sampled by the typical duration of a single continuous observation, requiring the use of different techniques, e.g. Miller et al. 2010; Zoghbi et al. 2013a.} (see also later studies, e.g. Kara et al. 2013; Chiang et al. 2017; Lobban et al. 2018; Mallick et al. 2018) and the relatively small scatter in the correlation with the BH mass, require significant contribution from the same optically thick reflecting/scattering layer to the soft excess.
This is predicted by photoionized disc relativistic reflection models, although, as suggested by Alston et al. (2014) and Gardner \& Done (2014), thermal reprocessing in a warm, optically thick inner corona may contribute to the observed soft lag in some sources.

\subsection{FeK lags}

Further evidence of inner disc X-ray reverberation came from the discovery of reverberation associated with the Fe K emission line. 
Fe K line reverberation is a major test for disc reflection models, as it occurs in a part of the spectrum which is not affected by other possible signatures of reprocessing (e.g. thermal reprocessing, Gardner \& Done 2014; warm absorbers, Silva et al. 2016).  

Fe K line reverberation was first reported by Zoghbi et al. (2012), from a study of the bright Seyfert 1.5 galaxy NGC 4151. This source shows shorter time delays ($\sim$1 ks) in the red wing of the line than in the narrow core ($\sim$2 ks) with respect to the X-ray continuum, suggesting the two components to be produced at different radii. The measured delays would then correspond to distances of $\sim$2-4 $r_g$ for a BH mass of $4.5\times10^7 M_{\odot}$ (Kara et al. 2016b) and assuming an X-ray source located above the disc. Fe K lags were later revealed in several other individual sources (e.g. Kara et al. 2013; Zoghbi et al. 2013b; Zoghbi et al. 2014).
Kara et al. (2016b) carried out a systematic study of Fe K reverberation lags in a sample of sources partially overlapping the sample analyzed in De Marco et al. (2013b), but including also more obscured AGN. They report detection in 20 sources. 
The measured high-frequency Fe K lags scale with BH mass in the same fashion as observed for soft lags and map similar light-crossing distances of a few $r_g$, as expected if due to the same reprocessing mechanism. While these results imply that reflection plays a role also in the production of the observed soft lags, any other possible form of reprocessing additionally contributing to the soft excess (e.g. Gardner \& Done 2014) should come from the same or very close reprocessing regions as those where the reflection spectrum is produced.

Alternative interpretations involving a large-scale reflector have also been proposed (e.g. Miller et al. 2011; Mizumoto et al. 2018). However, whether these models can offer a self-consistent explanation of the ubiquity of soft lags in radio quiet unobscured AGN, the detection of Fe K lags, and the more general spectral-timing properties of these sources is debated (see also discussion in Uttley et al. 2014). 

\section{X-ray reverberation in BHXRB}\label{sec:revBHXRB}

Being a diagnostic of the inner flow geometry, X-ray reverberation finds tantalizing applications in the study of BHXRBs. Indeed, theoretical models tell us that major changes of inner flow geometry as a function of the accretion state must occur to explain their spectral and timing evolution during an outburst (e.g. Esin et al. 1997; Poutanen et al. 1997; Meyer et al. 2000). 

\begin{figure}
        \includegraphics[width=\columnwidth]{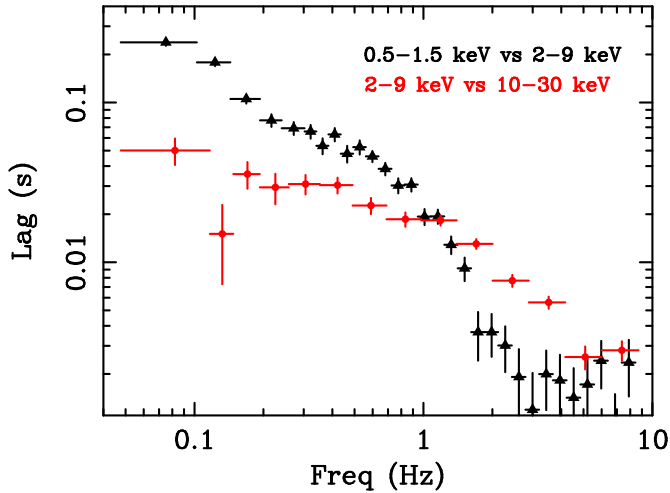}
    \caption{The lag vs frequency spectrum of GX 339-4 in a bright hard state ($L_{3-10keV}/L_{Edd} \sim$0.02) of the 2010/2011 outburst (from De Marco et al. 2015b).} 
    \label{fig:GXlagfreq}
\end{figure}

BHXRB have long been studied using Fourier timing techniques, with the first measurements of hard X-ray lags dating back to the 80's (e.g. Miyamoto et al. 1988). Hard lags (lags of the hard X-ray photons with respect to soft ones) are commonly observed in BHXRBs (Nowak et al. 1999; Pottschmidt et al. 2000; Grinberg et al. 2014; Fig. \ref{fig:GXlagfreq}). They are also observed at low frequencies in several AGN (De Marco et al. 2013; Fig. \ref{fig:1H}). These lags are intrinsic to the broad band X-ray continuum and they are commonly interpreted as due to perturbations in the mass accretion rate, propagating from larger to smaller radii of the accretion flow (e.g. Kotov et al. 2001).

Contrary to hard lags, evidence of X-ray reverberation in BHXRBs was only recently found (Uttley et al. 2011), and, to date, the number of detections is still significantly smaller than in AGN. This has to be ascribed to a number of reasons. In BHXRBs the light-crossing time of one $r_g$ of the source is much smaller than in
AGN, thus affecting the net number of collected photons on the time scales relevant for X-ray reverberation\footnote{However, the advantage of BHXRBs is that, beside getting brighter than AGN by about 3 orders of magnitude, the duration of a X-ray observation typically exceeds the X-ray reverberation light-crossing times by several orders of magnitude, thus allowing averaging over many variability cycles to improve the statistics.}. In addition, the sensitivity of lag measurements depends on the intrinsic variability power of the source, which in BHXRBs is higher during hard states (e.g. Mu\~noz-Darias et al. 2011). However, given its remarkable timing capabilities, studies of BHXRBs in the past were mostly carried out using data from the {\it Rossi X-ray Timing Explorer} (RXTE), whose sensitivity was limited to $E \gsim 2$ keV. This precluded direct observation of the disc during the hard state (as the inner disc temperature in the hard state is significantly lower than in the soft state, e.g. Tomsick et al. 2008), which later turned out to be crucial for detection of X-ray reverberation. 

Indeed, the first evidence of X-ray reverberation in a BHXRB came from the study of a long {\it XMM-Newton} observation of GX 339-4 during its hard state (Uttley et al. 2011). The use of data from the EPIC-pn instrument in its fast-readout mode allowed studying the lags down to $\sim$ 0.4 keV, thus revealing the disc in the hard state. Studying the high-frequency ($\nu\sim 2-8$ Hz) lags of the source, the authors found signatures of a soft component lagging behind the broad band hard X-ray continuum by a few milliseconds. This was interpreted as due to {\it thermal reverberation}: a fraction of the observed disc variability in the hard state (Wilkinson \& Uttley 2009; De Marco et al. 2015a) is driven by heating of the variable comptonized X-ray photons (e.g. Malzac et al. 2005), thus lagging behind the hard X-ray continuum. This component becomes visible at high-frequencies ($\nu \gsim 2$ Hz), where the hard lag amplitude is smaller (Fig. \ref{fig:GXlagfreq}). The measured thermal reverberation lag corresponded to a light travel time distance of a few tens of $r_g$ of the X-ray source to the reprocessing region. This lag is longer than expected by rescaling the amplitude of the lags observed in AGN (Fig. \ref{fig:AGNlagMass} right panel, and figure 9 in De Marco et al. 2013a), indicating a different inner flow geometry associated with the hard state.

\begin{figure*}
        \includegraphics[width=\columnwidth]{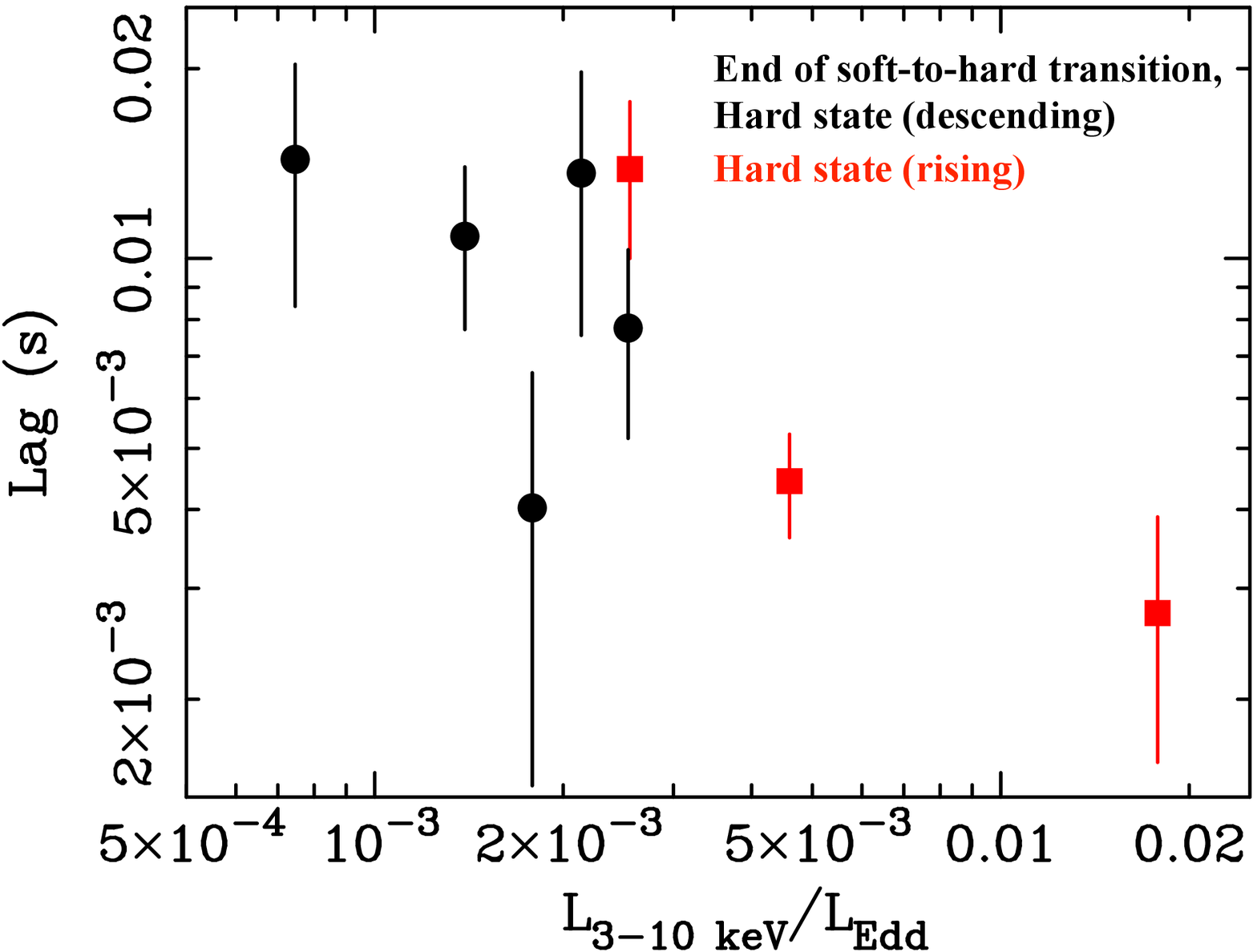}
        \hspace{0.2cm}
        \includegraphics[width=\columnwidth]{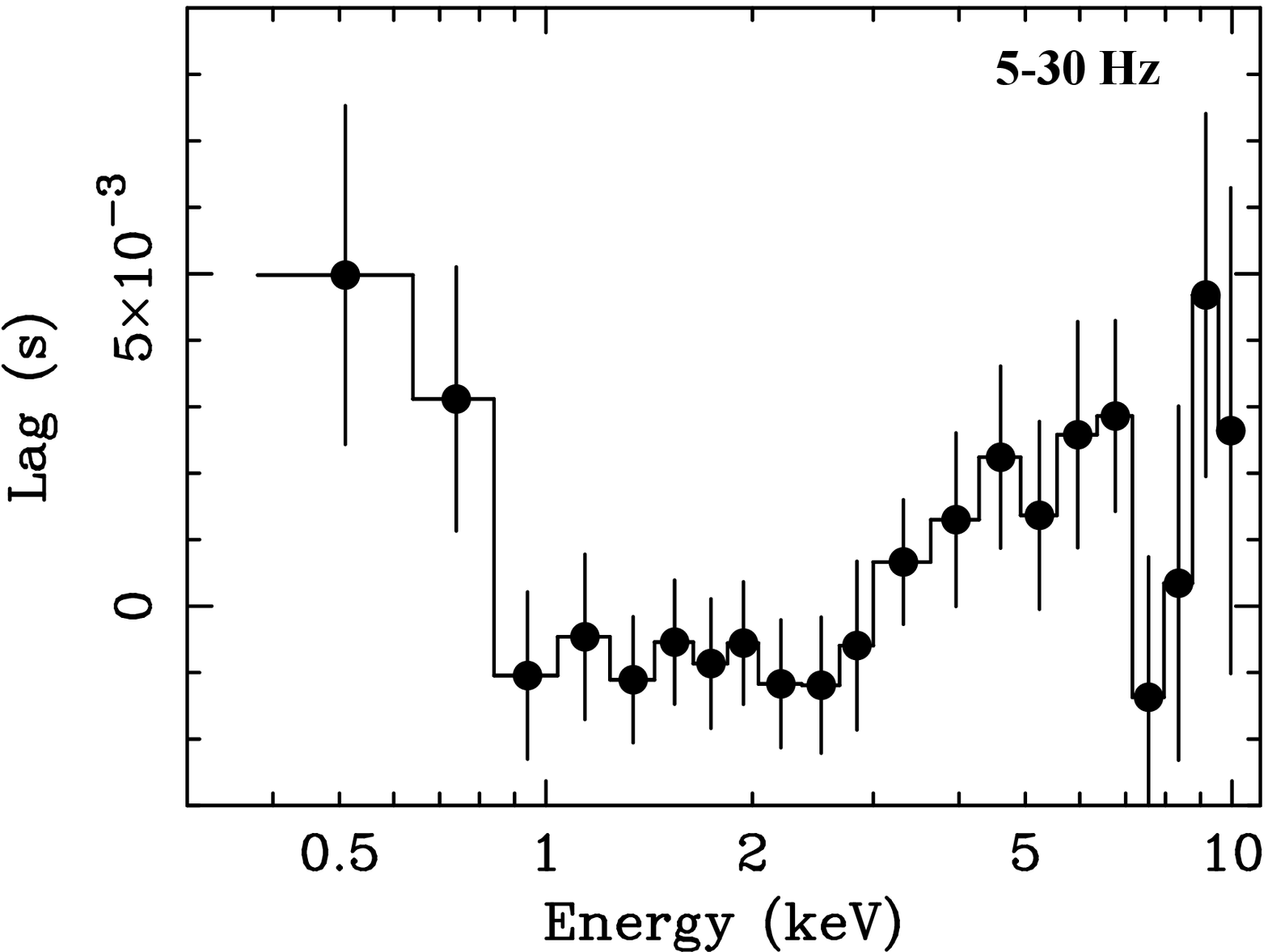}
    \caption{{\it Left :} Reverberation lag amplitude as a function of $3-10$ keV Eddington-scaled luminosity in the hard state of GX 339-4. {\it Right:} The lag vs energy spectrum of GX 339-4 in a hard state observation of the 2014-2015 outburst. The thermal reverberation lag is seen at $E\lsim1$ keV, signatures of Fe K line reverberation are seen at $\sim$ 6 keV (from De Marco et al. 2017).}
    \label{fig:GXlagvsL}
\end{figure*}

This discovery triggered us to systematically search for X-ray reverberation in a larger sample (De Marco et al. 2015b), with the goal to study the geometrical changes of the inner flow as a function of the accretion state. To this aim we analyzed available XMM-Newton observations of 10 BHXRBs. We found evidence of thermal X-ray reverberation in a total of 4 observations of GX 339-4 and 4 observations of H1743-322 (De Marco et al. 2015a; De Marco \& Ponti 2016) all during the hard state\footnote{The large number of non-detections can be ascribed to the fact that most of the available {\it XMM-Newton} observations were characterized by a short exposure and/or caught the source in a low fractional variability soft state or in a dim hard state.}. Interestingly, the detected reverberation lags are longer (by a factor $\sim$10-100) than inferred by rescaling those seen in AGN for the difference in BH mass and assuming the same geometry, but they show a trend of decreasing amplitude as a function of luminosity throughout the hard state (Fig. \ref{fig:GXlagvsL}). This is expected if the geometry of the inner accretion flow evolves. In particular, the trend indicates that the distance mapped by the thermal reverberation lag smoothly decreases as the source evolves throughout the first stages of the outburst.
A recent {\it XMM-Newton} campaign which monitored GX 339-4 during the last phases of the soft-to-hard state transition and the decrease of luminosity in the hard state preceding its return to quiescence confirmed this result (De Marco et al. 2017; Fig. \ref{fig:GXlagvsL}).

The observed scaling of the X-ray reverberation lag with luminosity follows predictions of truncated disc models (e.g. Esin et al. 1997, Meyer et al. 2000), whereby the inner disc truncation radius should approach the ISCO as the luminosity increases at the beginning of the outburst and then recede as the luminosity decreases at the end of the outburst. Interestingly, the fact that the reverberation lag amplitude is still quite large ($\sim$0.003 s) in the very bright hard state (i.e. at Eddington-scaled luminosities of $L_{3-10keV}/L_{Edd} \sim$0.02) might indicate substantial disc truncation throughout the entire hard state (but see Sect. \ref{sec:discussion}). Alternative proposed explanations involve a disc always reaching the ISCO, with the source of hard X-ray photons becoming more compact as the luminosity increases.

As for AGN, we expect Fe K reverberation to be observable also in BHXRBs. However, this requires good statistics at the highest frequencies, where the lags of the X-ray continuum are smaller and the lags associated with the discrete Fe K feature more easily distinguishable.  
Extending our lag measurements to the 5-30 Hz frequency range we could find some hint of reverberation in the FeK component during one observations of GX 339-4 (Fig. \ref{fig:GXlagvsL}, right panel). Higher quality data (e.g. from NASA's Neutron star Interior Composition ExploreR, NICER, Gendreau et al. 2012) are needed to allow us using FeK lags to constrain the evolution of inner flow geometry.

\section{Discussion}
\label{sec:discussion}

The use of X-ray spectral-timing techniques, in particular to measure X-ray reverberation, have opened a new window in the study of the accretion flow around accreting BHs. This kind of analysis requires long observations and/or high number of collected counts over short time scales. Therefore, a revolution is expected in this area with the advent of larger collecting area X-ray facilities, with fast-timing capabilities. These include the already operating NICER payload, sensitive in the same energy range as XMM, but with $\sim2\times$ higher effective area at 1.5 keV, $\sim100-1000\times$ better time resolution, and capable of observing very bright sources ($> 1$ Crab) with negligible dead time and pile up (Gendreau et al. 2012). 
In the future, eXTP (enhanced X-Ray Timing and Polarimetry; De Rosa et al. 2018) and ATHENA (Advanced Telescope for High ENergy Astrophysics; Dov\v{c}iak et al. 2013) are expected to increase the sensitivity of lag measurements by a factor up to $\sim$40 at 6keV and up to $\sim$10 at 1 keV, respectively.

However, observational studies must progress in tandem with theoretical ones. 
In particular, the development of X-ray spectral-timing models is crucial for obtaining more stringent constraints on the geometry of the inner accretion flow.
In the absence of such models, only back-of-the-envelope estimates can be derived. These models can account for several effects, including dilution (see Uttley et al. 2014), the multiple paths of photons from the X-ray source to the disc, the coronal geometry, and the behaviour of lags intrinsic to the primary X-ray continuum. Given the quality of the data analysed so far, the most significant effect on current measurements is probably the reduction of observed lag amplitude produced by dilution. Nonetheless, dilution does not affect the frequency at which the lag is suppressed (Uttley et al. 2014). Access to these high frequencies can be strongly limited by statistics. However, in bright AGN current measurements extending to those frequencies suggest small size scales to the reprocessing region ($\lsim 10 r_g$ ), meaning that dilution is not very important in these datasets. Notably, this purely observational inference is backed up by results from fits with currently available spectral-timing models (Cackett et al. 2014; Emmanoulopoulos et al. 2014; Epitropakis et al. 2016; Chainakun et al. 2016). 

For BHXRBs, on the other hand, larger effective area X-ray detectors will be decisive to access the high-frequency regime. 
The existence of an evolution of the reverberation lag as a function of luminosity in the hard state of BHXRBs is an indication of disc truncation (De Marco et al. 2015b; 2017). However, a precise estimate of the truncation radius requires spectral-timing models fitting. This implies proper modelization of the continuum hard lags, as the latter are still significant in the frequency range of current detections. 
Several groups have been independently studying this problem, from an analytical (Ingram \& van der Klis 2013; Rapisarda et al. 2016; 2017; Mahmoud \& Done 2018; Mahmoud, Done \& De Marco 2018; Mastroserio et al. 2018) and numerical (Ar\'evalo \& Uttley 2006; Ingram \& Done 2012) point of view, as well as in the context of magnetohydrodynamic (MHD) simulations (Hogg \& Reynolds 2016), so that the number of available models is rapidly increasing.

\section{Summary}

X-ray reverberation is an independent method to investigate the geometry of the emitting regions around accreting BHs. 
X-ray reverberation lags are now commonly observed in radio quiet AGN. The mapped light-crossing distances and the scaling with BH mass are indicative of similar inner flow geometry in these sources. The measured lag amplitudes and characteristic frequencies are consistent with the presence of a compact corona and a disc extending down to the ISCO. X-ray reverberation lags in BHXRBs show a trend of decreasing lag amplitude as a function of luminosity in the hard state. This behaviour is suggestive  of variations of inner flow geometry and is in agreement with predictions of truncated-disc models.


\section*{Acknowledgments}

BDM acknowledges the \fundingAgency{Polish National Science Center} (Polonez \fundingNumber{2016/21/P/ST9/04025}). GP acknowledges the Bundesministerium f\"{u}r Wirtschaft und Technologie/Deutsches Zentrum f\"{u}r Luft und Raumfahrt (BMWI/DLR, FKZ 50 OR 1812, OR 1715 and OR 1604) and the Max Planck Society. %










\bibliography{Wiley-ASNA}%



\end{document}